\documentclass[aps,prl,twocolumn,showpacs,superscriptaddress,floatfix,10pt]{revtex4-1}
\usepackage{amsmath,amssymb,amsfonts,stmaryrd,wasysym,graphicx,multirow,color,textcomp,subfigure}
\usepackage{url}
\usepackage[colorlinks=true,citecolor=blue,urlcolor=blue]{hyperref}
\usepackage{romannum}
\usepackage{color,soul} 
\usepackage{bm}
\usepackage[normalem]{ulem}

\hypersetup{colorlinks=true, urlcolor=blue, citecolor=blue, pdfborder={0 0 0}}

\normalfont
\def\be{\begin{equation}}
\def\ee{\end{equation}}
\def\bea{\begin{eqnarray}}
\def\eea{\end{eqnarray}}

\begin{document}
\title{Higher-Order Topological Instanton Tunneling Oscillation}
\author{Moon Jip Park}
\email{moonjippark@kaist.ac.kr}
\affiliation{Department of Physics, KAIST, Daejeon 34141, Republic of Korea}

\author{Sunam Jeon}
\email{sunamfx@naver.com}
\affiliation{Department of Energy Science, Sungkyunkwan University, Suwon 16419, Korea}

\author{SungBin Lee}
\email{sungbin@kaist.ac.kr}
\affiliation{Department of Physics, KAIST, Daejeon 34141, Republic of Korea}

\author{Hee Chul Park}
\email{hcpark@ibs.re.kr}
\affiliation{Center for Theoretical Physics of Complex Systems, Institute for Basic Science (IBS), Daejeon 34126, Republic of Korea}

\author{Youngkuk Kim}
\email{youngkuk@skku.edu}
\affiliation{Department of Physics, Sungkyunkwan University, Suwon 16419, Korea}

\date{\today}

\begin{abstract}	
We propose a new type of instanton interference effect in two-dimensional higher-order topological insulators. The intercorner tunneling consists of the instanton and the anti-instanton pairs that travel through the boundary of the higher-order topological insulator. The Berry phase difference between the instanton pairs causes the interference of the tunneling. This topological effect leads to the gate-tunable oscillation of the energy splitting between the corner states, where the oscillatory nodes signal the perfect suppression of the tunneling. We suggest this phenomenon as a unique feature of the topological corner states that differentiate from trivial bound states.  In the view of experimental realization, we exemplify twisted bilayer graphene, as a promising candidate of a two-dimensional higher-order topological insulator.  The oscillation can be readily observed through the transport experiment that we propose. Thus, our work provides a feasible route to identify higher-order topological materials.
\end{abstract}

\maketitle
\textbf{Introduction-} The exploration of the defects and the bound states is one of the main themes in condensed matter physics. Along this line, the recent theoretical discovery of the higher-order topological insulator(HOTI) has drawn a great attention for the realization of a new type of zero-dimensional bound state, known as the topological corner states\cite{Benalcazar61,PhysRevLett.110.046404,PhysRevB.96.245115,PhysRevLett.119.246401,PhysRevLett.119.246402,Schindlereaat0346,PhysRevLett.120.026801,Serra-Garcia2018,Peterson2018,Imhof2018,PhysRevB.97.205136,PhysRevB.97.205135,PhysRevX.8.031070,PhysRevB.97.241405,PhysRevLett.121.096803,PhysRevB.98.045125,Xue2019,PhysRevB.98.081110,PhysRevX.9.011012,PhysRevLett.121.116801,PhysRevB.97.041106,PhysRevB.98.241103,PhysRevLett.122.076801,PhysRevB.98.201402,PhysRevB.99.020304,PhysRevB.99.085406,PhysRevB.98.201114,PhysRevB.98.245102,wheeler2018many,kang2018many,ono2019difficulties,wieder2018axion,agarwala2019higher,PhysRevB.99.041301}. The topological corner states emerge as the physical manifestation of the quantized quadrupole moment at the corner of the two-dimensional HOTI. The corner states stand in stark contrast to trivial bound states in that they are embedded in the boundary of the higher-dimensional manifold. There are now several theoretical proposals for the material candidates of the two-dimensional HOTI. The examples in atomic solids include the twisted bilayer graphene\cite{PhysRevX.9.021013,PhysRevLett.114.226802,PhysRevLett.123.216803,PhysRevLett.123.036401}, phosphorene\cite{PhysRevB.98.045125}, and monolayer graphdiyne\cite{doi:10.1021/acs.nanolett.9b02719,lee2019higher,sheng2019two}. The area of HOTI also extends toward a variety of wave phenomena in metamaterials including photonics\cite{PhysRevB.98.205147,PhysRevLett.122.233903,Mittal2019} and acoustics\cite{Xue2019,Ni2019,Zhang2019,PhysRevB.101.161301}. 

\begin{figure*}[t!]
	\centering\includegraphics[width=0.85\textwidth]{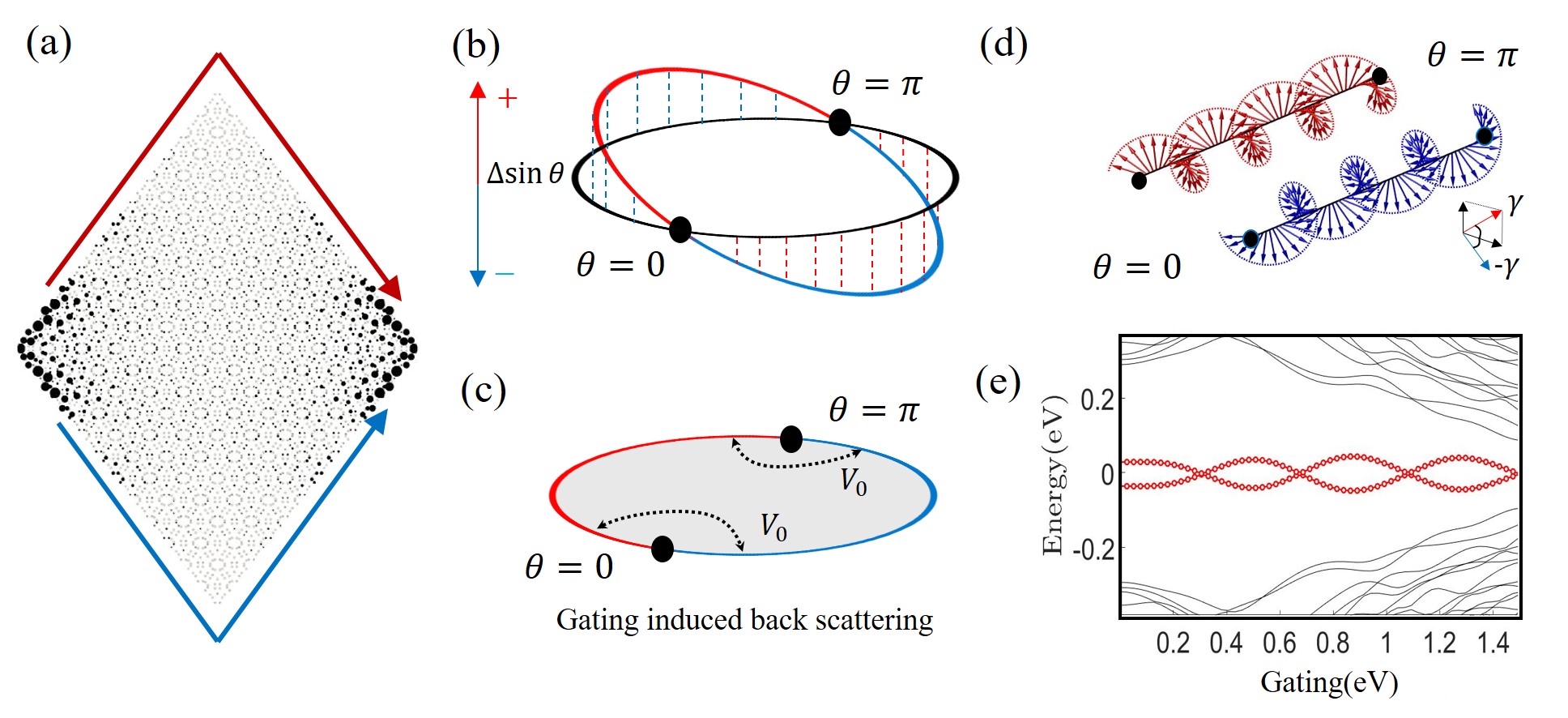}
	\caption{
		(a) The wave function distribution(black dots) of the corner states in TBG. Red and blue arrow illustrate the instanton tunneling paths that connect the corner states. The two paths as a pair fully circle the edge of the HOTI. 
		(b) Schematic figure illustrating the Jackiw-Rebbi construction of the HOTI. The helical edge modes are localized in the disk geometry. We introduce an angular dependent hybridization between the edge modes(The  red and blue lines indicates the sign change of the hybridization). The hybridization gap vanishes at $\theta=0$ and $\theta=\pi$, which form a domain wall, where the corner states are localized. 
		(c) The mirror symmetry relates the two instantons and there is no phase difference between the two. However, the application of the gate voltage introduce the additional backscattering term which breaks the mirror symmetry. This backscattering term can be thought as the effective flux, which change the sign depending on the paths.
		(d) The two instanton paths acquire the opposite Berry phases(red and blue arrows), $\pm\gamma$, as the electron travels from $\theta=0$ to $\theta=\pi$. This leads to the instanton interference effect between the instantons.
        (e) Energy spectrum of the HOTI as a function of a gate voltage. As the two instantons interfere each other, the intercorner tunneling amplitude oscillates as a function of the gate voltage.	
}
	\label{Fig1}
\end{figure*}

The rapid progress in the higher-order topological matters calls for the understanding of new experimental properties with potential applications in technology. In this letter, we study the novel quantum tunneling effect between the corner states. Unlike the conventional double-well tunneling problem, the corner states are connected by the instanton paths that circulate the closed edge of the two-dimensional bulk of HOTI. An instanton path and its complementary anti-instanton path as a pair form a full circle of the edge[See Fig.\,\ref{Fig1} (a)]. Our main discovery is the quantum interference effect between the instanton paths, arising from the intrinsic Berry phase as the electron adiabatically circulates the HOTI bulk. This topological interference effect manifests as the oscillatory behavior of the intercorner tunneling and the energy splitting of the corner states, $\Delta E $, as shown in Fig.\,\ref{Fig1} (e). In particular, the oscillatory nodes are characterized by the perfect suppression of the tunneling, signaling the destructive interference between the instantons and the anti-instantons. 

Another important aspect of this paper is to propose a feasible experimental platform of the instanton interference effect in realizable conditions. In this regard, we exemplify the case of twisted bilayer graphene (TBG), which has been proposed as a promising candidate for the two-dimensional HOTI\cite{PhysRevLett.123.216803}. We demonstrate that the instanton tunneling oscillation is measurable through resonant transport. As a result, our work offers a promising platform for the hunt for the two-dimensional HOTI.

\textbf{Higher-order Jackiw-Rebbi Soliton-} We start our discussion by presenting a simple model capturing the higher-order topology of TBG. TBG is a representative model of the two-dimensional HOTI protected by the second Stiefel-Whitney(2nd SW) number in the presence of the space time-reversal symmetry($P\times T$) regardless of the twist angle \footnote{For the detailed implementation of the higher-order topology, see supplementary \ref{sec:HOTI} and Ref. \cite{PhysRevLett.123.216803,PhysRevLett.114.226802,RevModPhys.88.035005,ma2019discovery}}. The origin of the corner states can be intuitively understood using the equivalent $\mathbb{Z}_2$ mirror winding number in the presence of the additional mirror symmetry. The $\mathbb{Z}_2$ mirror winding number can be defined as the mirror projected Zak phase, $\nu_\pm$, along the mirror invariant line. Here, $\pm$ indicates the subsectors characterized by the mirror eigenvalue, $M=\pm 1$. The non-trivial mirror winding number physically manifests as the emergence of the 1D counter-propagating edge modes only if the boundary termination is mirror-symmetric. 

In general, the boundary termination may not preserve the mirror symmetry. In such case, the global structure of the edge mode can be described by the Jackiw-Rebbi soliton formulation of the HOTI\cite{PhysRevLett.123.186401,Wieder2020}. The formulation considers the edge modes in the disk geometry of the radius, $R$, with the angle dependent hybridization[See Fig.\,\ref{Fig1} (b)]:
\bea
\nonumber
H(\theta)= \Psi^\dagger(\theta)[ i\frac{v_F}{R}\partial_\theta \tau^z+\Delta \sin \theta(e^{i\theta}\tau^++e^{-i\theta}\tau^-) ]\Psi(\theta),
\\
\label{eq:H}
\eea
where $\Psi(\theta)=(\psi_1(\theta),\psi_2(\theta))^T$ is the spinor of counter propagating edge mode with the Fermi velocity $v_F$. $\tau^z=\pm 1$ represents the chirality of the edge modes. $\Delta \sin\theta$ represents the angle dependent hybridization. The strength of the hybridization vanishes at the mirror symmetric corners($\theta=0,\pi$) that host an effective domain wall of the well-known Su-Schrieffer-Heeger(SSH) chain\cite{PhysRevLett.42.1698}. The physical consequence is the emergence of the localized zero modes at $\theta=0$ and $\theta=\pi$, known as the topological corner states. In addition, the Hamiltonian in Eq. \eqref{eq:H} preserves the space time-reversal symmetry, defined as, $P\times T: H(\theta)=\tau^x H^*(\theta+\pi) \tau^x$, in addition to the mirror symmetry, which is defined as $M: H(\theta)=\tau^y H(-\theta) \tau^y$  \cite{footnote1}.

\textbf{Instanton Oscillation in HOTI-} We now consider the intercorner tunneling, which lifts the degeneracy of the corner states. Two independent tunneling paths connect the corner states[See Fig.\,\ref{Fig1} (c)]. One is the clockwise circulations, and the other is the counter-clockwise circulations. The mirror symmetry, when it is preserved, relates the two paths, and thus the instanton representing these paths do not acquire phase difference. However, if the external gate voltage, which energetically splits the top and the bottom layers, is applied, the mirror symmetry is explicitly broken, while the space time-reversal symmetry, ensuring $e/2$ corner charge, is still intact. Once the mirror symmetry is broken, the finite phase difference between the two instantons can enter, and it leads to the interference effect.

Before presenting the numerical results with the full bulk lattice model, we introduce a heuristic analysis illustrating the instanton interference effect. The gate voltage, $V_0$, introduces a short-ranged backscattering to Eq. \eqref{eq:H}:
$
 H_{\textrm{gate}}(\theta)=V_0f(\theta)\Psi^\dagger(\theta) i\tau^y \Psi(-\theta),
 \label{eq:gating}
 $
 where $f(\theta)$ represent the angle dependent scattering between the edge modes \cite{footnote1}. We find that, in respect of the edge mode basis, $c_\pm(\theta)=\frac{1}{\sqrt{2}}[\psi_1(\theta)\pm\psi_2(-\theta),\psi_2(\theta)\pm\psi_1(-\theta)]$, the projected Hamiltonian becomes $\sum_{k}c^\dagger_\pm[i\frac{v_F}{R}\partial_\theta \pm V_0f(\theta)]\sigma_z c_\pm$ in the limit where $\Delta=0$. In other words, the gate voltage acts as the effective flux, similar to that of the Aharonov-Bohm effect\cite{PhysRevLett.5.3,PhysRevLett.56.792,PhysRevA.34.815}. Yet, an important difference with the Aharonov-Bohm effect is that the gate voltage acts as a pseudo-flux that preserves the time-reversal symmetry.
 
 The overall effect of the gate voltage is now to separate the two instanton paths by the opposite geometric phases, $\pm\gamma$, respectively, as the electron travel from $\theta=0$ to $\theta=\pi$. The phase $\gamma$ is explicitly evaluated as,
\bea
\gamma=\frac{2V_0R}{v_F}\int^\pi_0 d\theta f(\theta).
\label{eq:gamma}
\eea 
The direct physical manifestation of this novel interference effect is the oscillatory behavior of the energy splitting between the corner states: 
\bea
\Delta E^2 = 4K \sqrt{\frac{S_0}{2\pi }}e^{-S_0}| \cos(\gamma)|.
\label{eq:dE}
\eea
Here $S_0$ is the action of the instanton. $K$ is the constant determinant, describing fluctuations from the saddle point\cite{footnote1}. 
The energy splitting oscillates as a function of the geometric phase difference $\gamma$, and it vanishes whenever the destructive interferences occur($\gamma=(N+\frac{1}{2})\pi$, where $N\in \mathbb{Z}$). This feature establishes the gate-tunable instanton interference effect, which is the main finding of our paper. It is important to note that this result is based on the global gauge structure arising from the intrinsic Berry phase, and do not depend on the details of the wave functions. 

\begin{figure}[t!]
	\centering\includegraphics[width=0.5\textwidth]{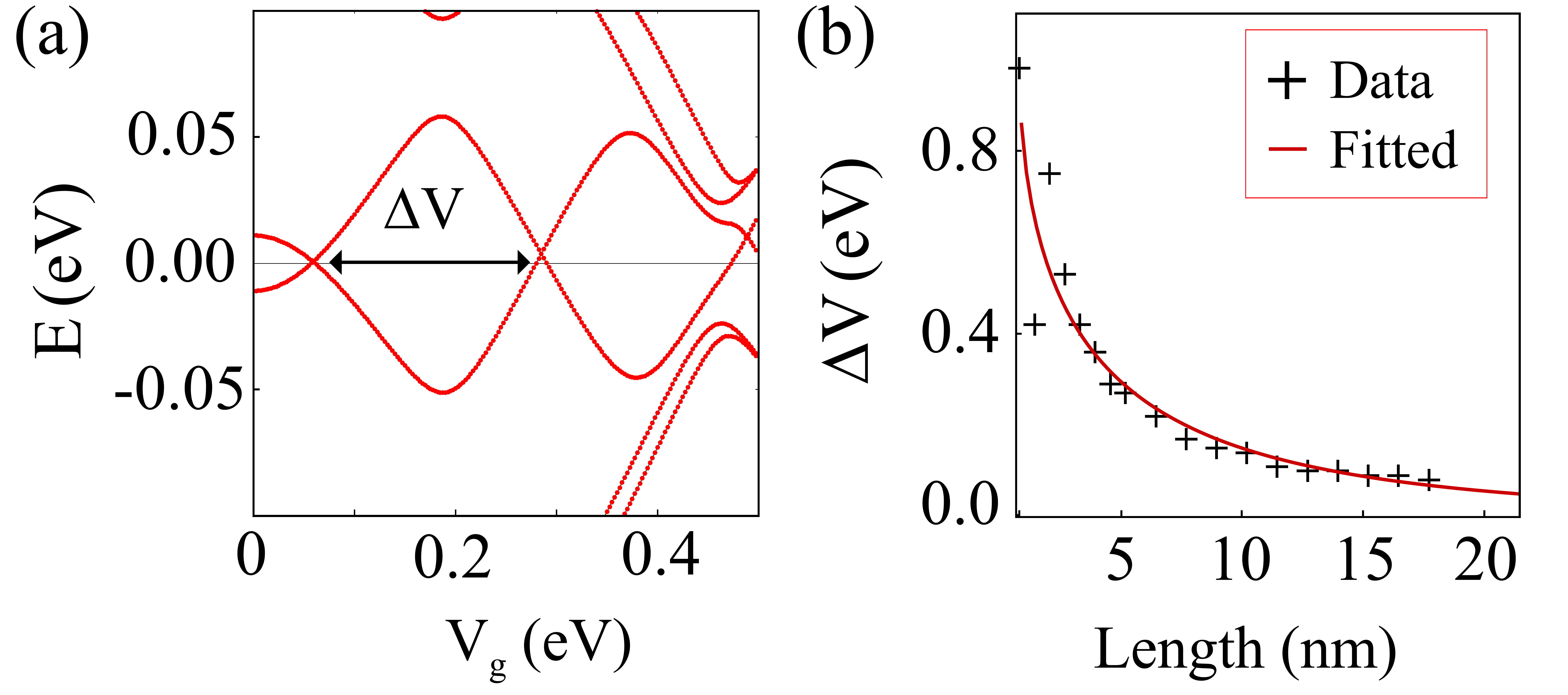}
	\caption{ 
		(a) The evolution of corner states energies as a function of the gate voltage, derived using the tight binding model of the TBG with the twist angle $\theta=22.79^\circ$. The energy splitting oscillates as a function of the gate voltage with the frequency $\Delta V$.
		(b) The frequency $\Delta V$ as a function of the width of the TBG.  
	}
	\label{Fig2}
\end{figure}

To estimate the realistic energy scale of the oscillation, we now utilize the finite-sized tight-binding model, using the set of parameters obtained from first-principles calculations \cite{PhysRevLett.123.216803}. We calculate the energy splitting between the corner states as a function of the gate voltage and the system size. Fig.\,\ref{Fig2} (a) shows the typical oscillation of the energy splitting between the corner states. Whenever the effective flux matches the commensurate values of the destructive interference, Fig.\,\ref{Fig2} (a) show that the energy splitting of the corner states exactly vanishes. The frequency of the oscillation is proportional to the system size as expected from Eq. \ref{eq:dE} and explicitly confirmed in Fig.\,\ref{Fig2} (b). As the width of the nanoribbon reaches $\sim140$ nm scale, we find that the oscillation of the instanton splitting requires the gate voltage of $150$ $\mu$V, which should be within the observable regime from the currently existing experimental techniques\cite{Mahapatra2017}.

\begin{figure*}[t!]
	\centering\includegraphics[width=1\textwidth]{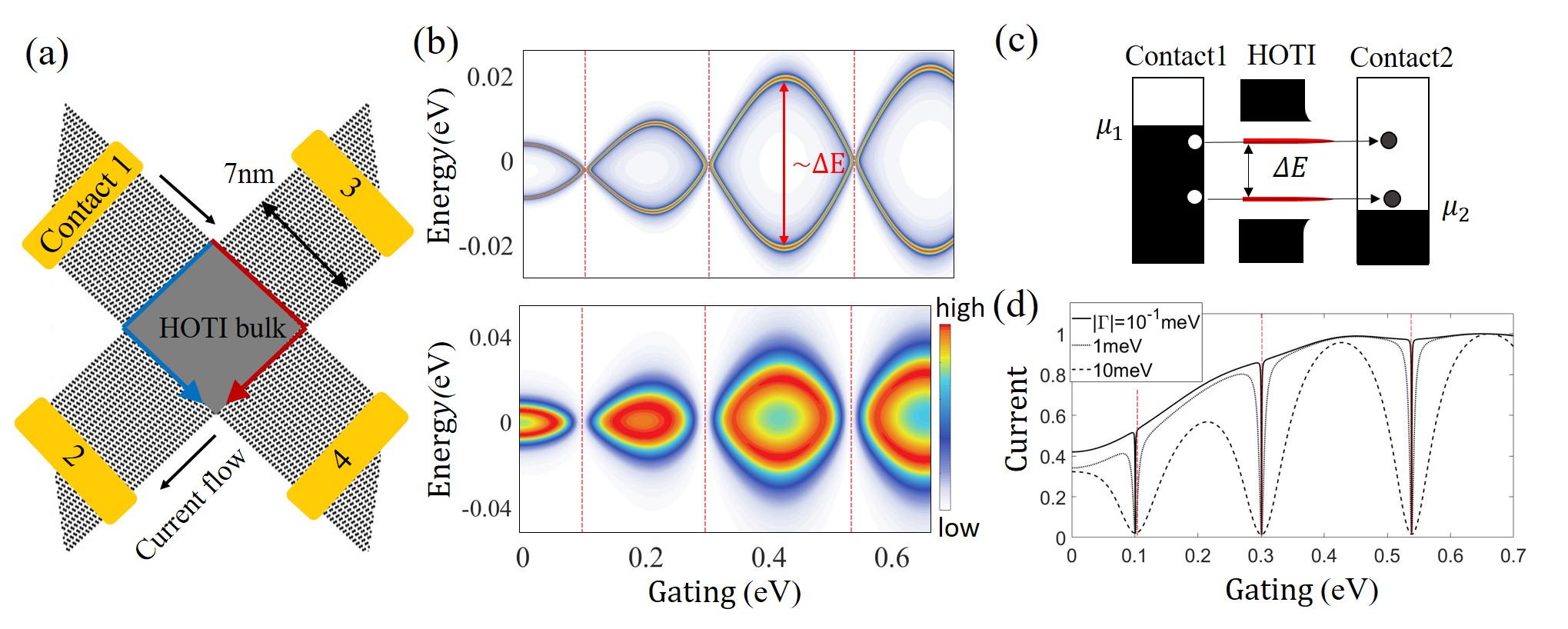}
	\caption{ (a) Transport setup of TBG, which consists of two misaligned graphene nanoribbons. The overlapping region of two nanoribbons forms the HOTI phase while the non-overlapping regions serve as the metallic contacts. The instanton tunneling between the corner states mediates the electronic transmission from contact 1 to contact 2. (b) Normalized transmission of electrons from contact 1 to contact 2 as a function of the gate voltage with various values of the broadening, (top) $|\Gamma|=1$meV and (bottom) $|\Gamma|=10$meV respectively. The non-zero transmission occurs through the instanton tunneling between the topological corner states. The instanton interference effect causes the oscillatory behaviors in the transmission peak. At the nodes of the oscillations, the transmission peaks vanish as they destructive interfere. (c) Schematic illustration of the Landauer-B\"uttiker transport in the proposed setup. Electron tunneling from contact 1 to contact 2 occur through the resonant tunneling of the corner states(Red line), separated by the instanton gap, $\Delta E$. (d) Normalized tunneling current derived by integrating the transmission in (b). The perfect suppression of the tunneling current occur at the nodes of the instanton oscillations(red lines).}
	\label{Fig3}
\end{figure*}

\textbf{Transport signature in TBG}- We propose that this instanton interference effect manifests as a well-marked transport signature of HOTI. Our proposed experimental setup comprises crossing two graphene nanoribbons as shown in Fig.\,\ref{Fig3} (a). The overlapping region forms the HOTI phase while the non-overlapping regions serve as the natural contact where the electrons are injected and collected. The core idea of this transport setup is that the electrons departed from the contact 1 need to undergo the instanton tunneling to reach the contact 2. To theoretically investigate the transport behavior, we derive the transmission from the contact 1 to contact 2, $T_{12}$, using the Landauer-B{\"u}ttiker formula\cite{datta_2005,Hirsbrunner_2019},
\bea
T_{12}=\textrm{Tr}[G \Gamma_1 G^\dagger \Gamma_2].
\eea 
Here, $G$ represents the retarded Green function of the HOTI region. $\Gamma_{n} \equiv i(\Sigma_{n}-\Sigma_{n}^\dagger)$ represents self-energy of n-th contact. Since the gate voltage is negligibly smaller than the band width of the graphene, Wide band limit(WBL) is employed to model the contact self-energy, $\Sigma_i=-\frac{i}{2}|\Gamma|$\cite{doi:10.1063/1.4793259}. For the transport simulation, we have used the enhanced band gap due to the size limit. However, the qualitative results will not change. 

The core result of the transport simulation is shown in Fig.\,\ref{Fig3}(b) by plotting the transmission as a function of the gate voltage and the energy. We first discuss the transport in the zero gate voltage. We find that the formation of the transmission gap, originating from the bulk gap of the HOTI. Inside the gap where the transmission is suppressed, we find the two in-gap resonant tunneling peaks, which originate from the corner states. The two peaks are separated by the energy scale of the corner splitting $\Delta E$. This separation indicates the intercorner instanton tunneling, which we discuss its behavior in detail. 

As the non-zero gate potential enters, the separation of the transmission peaks oscillates, which is the direct manifestation of the instanton interference effect we discussed in the previous section. This oscillation of the transmission peaks as a function of a gate potential is unexpected from trivial localized state, as the energy of the trivially localized states runs up (down) from the in-gap region as the electronic potential increases (decreases). 

More striking behavior of the HOTI instanton tunneling is the complete suppression of the transmission peaks whenever they merge each other at the nodes of the oscillations(colored by the red-dotted lines in Fig.\,\ref{Fig3} (b)). This suppression of the transmission is in stark contrast with the standard phenomenology of the conventional tunneling behavior. It is the consequence of the destructive interference followed by the $\pi$-Berry phase difference between the instantons. We would like to note that the observation of the destructive interference can be achieved from the measurement of the tunneling current, and it would not require a fine-tuning of the bias voltage as long as the chemical potentials are placed inside the in-gap regions(Fig.\,\ref{Fig3} (c)). In such case, the destructive interference of the instanton tunneling realizes as the nodes of the total tunneling current(red lines in Fig.\,\ref{Fig3} (d)). We find that the destructive interference patterns are well-survived even when the band broadening is comparable to the corner states splitting.

\textbf{Conclusion-}
In conclusion, we have studied the instanton tunneling between the corner states in the two-dimensional HOTI phase. Especially, we exemplified the case of the TBG, which represents the class of the HOTI phase protected by the second Stiefel-Whitney number. Unlike the conventional double-well tunneling problem, the instanton paths connecting the corner states pass through the edge of the HOTI. This holographic feature of the intercorner tunneling generates the intrinsic Berry phase differences between the different instanton paths, resulting in the novel instanton interference effect. The physical manifestation of the interference effect is the gate-tunable oscillatory pattern of the intercorner energy splitting. 

Focusing on this phenomenon, we have proposed the transport experiment to detect the two-dimensional HOTI phase. The node of the oscillation indicates the complete decoupling between the corner states, which would not be generally possible unless the corner states have the higher-order topological origin. In the tunneling transport experiment, the interference effect realizes as the suppression of the tunneling current. In our experimental setup, each graphene sheet serves as the natural contact, therefore we do not expect any ambiguity arising from edge termination or contact. In addition, we also expect the transport signature to be robust in the presence of the disorder. In general, there can be impurity states localized in the top or the bottom layer. However, the energy level of the impurity states will be monotonically biased under the application of the displacement field. One can readily separate the signature of the impurity from the corner states. 

\acknowledgements
M.J.P. thanks Hee Seung Kim for helpful discussions.
M.J.P. and S.L. are supported by the KAIST startup, the National Research Foundation Grant (NRF-2017R1A2B4008097) and BK21 plus program, KAIST. Y.K. acknowledges the support from the NRF Grant (NRF-2019R1F1A1055205). This work was supported by IBS through Project Code (IBS-R024-D1)  The computational resource was provided from the Korea Institute of Science and Technology Information (KISTI) (KSC-2019-CRE-0164). 
\bibliography{reference}

\clearpage
\pagebreak

\renewcommand{\thesection}{\arabic{section}}
\setcounter{section}{0}
\renewcommand{\thefigure}{S\arabic{figure}}
\setcounter{figure}{0}
\renewcommand{\theequation}{S\arabic{equation}}
\setcounter{equation}{0}

\begin{widetext}

\tableofcontents
\label{sec:HOTI}
\section{Higher-order topology in TBG}

We start our discussion by reviewing the crystalline symmetries and the associated bulk topology of twisted bilayer graphene(TBG)\cite{PhysRevLett.123.216803}. We construct the atomic configuration of TBG by twisting AA-stacked bilayer graphene at the hexagonal center with a give twist angle, $\theta$. Such twist preserves both $C_{6z}$ and $C_{2x}$ about the out-of-plane $z$- and in-plane $x$-axes, respectively, regardless of the specfiic angle $\theta$(See Fig. \ref{Fig1}). As a result, the crystalline symmetry of TBG belongs to the hexagonal space group \# 177 (point group  $D_6$). In contrast, the formation of moir\'{e} superlattice modifies the translational symmetry, depending on the twist angle. For any coprime integers $p$ and $q$,  a twist by $\theta_{p,q}\!=\!\arccos \frac{3p^2+3pq+q^2/2}{3p^2+3pq+q^2}$ results in an enlarged moir\'{e} unit cell with the lattice constant $L \!=\! a\sqrt{\frac{3p^2+3pq+q^2}{{\rm gcd}(q,3)}}$, where $a$ is the original lattice constant and gcd means the greatest common divisor\cite{PhysRevB.98.085435,PhysRevLett.99.256802}.

In the limit where $\theta \! \lesssim \!1^\circ$ without the lattice distortions, the Moir\'{e} potential has long periodicity in real space, resulting in negligible interaction between valleys.  In this limit, including the so-called {\em magic angles} where the Fermi velocity vanishes\cite{Bistritzer12233,PhysRevB.86.155449}, each valley is effectively decoupled and the  $U(1)$ valley symmetry $U(1)_v$ is approximately preserved and it, together with $C_{2z}\mathcal T$ symmetry, provides topological protection of four Dirac points associated with the $\mathbb{Z}_2$-quantized Berry phases $\pi$\cite{PhysRevLett.118.156401, 1807.10676}. Here, $\mathcal T$ represents time-reversal symmetry, where $\mathcal T^2 = 1$ without spin-orbit coupling. However, in generic angles, the $U(1)_v$ symmetry is not exact anymore and the fourfold-degenerate Dirac points can split into two pairs of massive Dirac points. Previous studies have reported the presence of the global gap at large angles. The intervalley coupling, and thus the gap opening between Dirac points, has a tendency to increase as the twist angle $\theta$ increases\cite{PhysRevB.81.165105,PhysRevLett.101.056803}.

\begin{figure}[t!]
	\centering\includegraphics[width=0.8\textwidth]{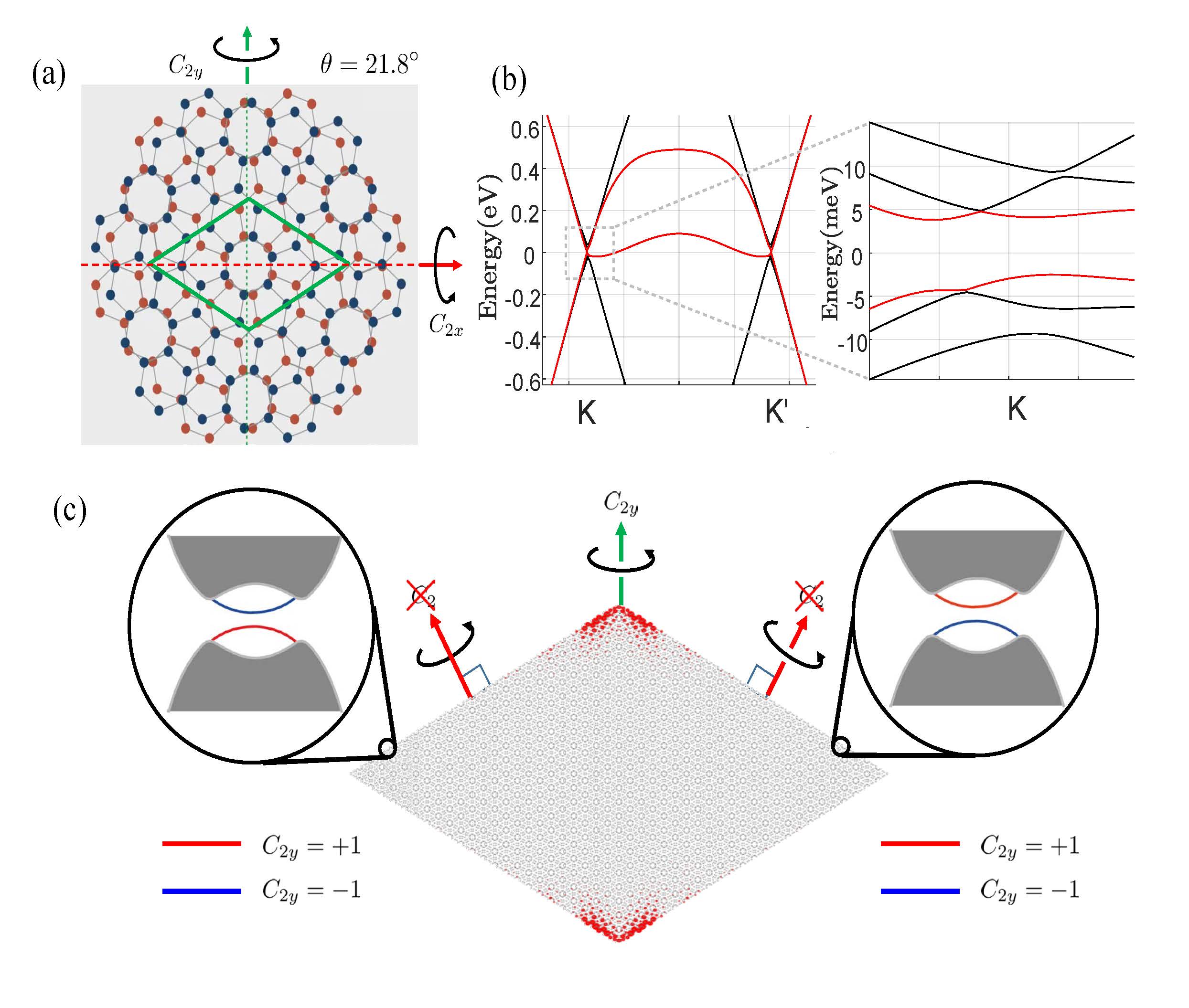}
	\caption{(a) Schematic illustration of the atomic structure of moir\'{e} unit cell of TBG at the twist angle $\theta = 21.8^\circ$. (b) Illustration of the edge spectrum. The green box indicates the moir\'{e} unit cell. The red (blue) circles represent the carbon atoms in the upper (lower) layer. The boundary cut breaks $C_{2x}$ symmetry, in which the helical edge modes can backscatter each other and open a mini bandgap. Consequently, the edge spectrum generates a gap. (c) Illustration of the HOTI phase in the TBG flake. The left and the right edge has the opposite sign of the effective edge gap, which forms a 1D domain wall. On the corner where the gap closes, $e/2$ corner charge occurs.}
	\label{fig2}
\end{figure}

A notable property of the TBG, also applicable to generic honeycomb lattices, is that all moir\'{e} superlattice forms such that its size is an odd integer multiple of the pristine lattice ($L^2/a^2\equiv 2N+1$). Correspondingly, the moir\'{e} BZ in momentum space folds the monolayer BZ $2N+1$ times. While $2N$ mini-BZs (out of $2N+1$) form inversion pairs each other,  the remaining unpaired mini-BZ can contribute unpaired odd parity to the parity structure of the occupied bands. Surprisingly, this unique parity structure dictates the inversion symmetric HOTI phase characterized by the second Stiefel-Whitney (SW) number $\omega_2$ \cite{PhysRevLett.121.106403, Ahn_2019}:
\begin{align}
(-1)^{\omega_2} =  \prod_{\Gamma_i\in{\rm TRIM} } (-1)^{[N_{\rm occ}^-(\Gamma_i)/2]},
\label{eq:SW}
\end{align}
where $N^-_{\rm occ}$ counts the number of parity-odd occupied bands at a time-reversal invariant momentum (TRIM) $\Gamma_i$. As a result, all moir\'{e} structure of TBG with a bandgap hosts non-trivial higher-order topology, characterized by second SW number, regardless of $N$, and thus, the specific twist angle or the microscopic details of the atomic structure. The physical consequence of the non-trivial second SW number is the filling anomaly, which ensures localized $e/2$ charge at each corner at the half-filling\cite{benalcazar2018quantization}.

\subsection{Mirror Winding number} 
Besides the second SW number, there exists another topological invariant in  TBG, which is a mirror winding number\cite{PhysRevLett.114.226802,RevModPhys.88.035005}. Along the mirror invariant line in the BZ, the Bloch Hamiltonian can be decomposed into two distinct sub-sectors characterized by the mirror eigenvalues $C_{2x}=\pm 1$. The $\mathbb{Z}_2$ mirror winding number is defined as the mirror-resolved Zak phase, $\nu_\pm$ for $C_{2x}=\pm 1$ subsector:
$\nu_\pm = \frac{1}{i\pi}\log\det[\mathcal{U}_\pm],$
where $\mathcal{U}_\pm$ indicates the mirror projected Wilson line. The mirror winding number has $\mathbb{Z}_2$ classification and, in principle, it is independent of the second SW number. However, in the case where an additional $C_{3z}$ symmetry is present, the second SW number and the mirror winding numbers are formally equivalent\cite{PhysRevLett.123.216803}.

The emergence of the topological corner state is readily seen by utilizing the mirror winding number. The physical manifestation of the mirror winding number is the 1D helical edge modes on the mirror-symmetric boundary termination. In this case, each mirror sector $C_{2x}=\pm1$ carries counter-propagating chiral edge modes of one another. However, a generic edge termination is incompatible with the $C_{2x}$ symmetry. The corresponding edge spectrum gains an energy gap, which has correspondence with Eq. \eqref{Eq:inst}. Fig.\,\ref{fig2} (b) explicitly shows the gapped edge spectrum with the diagonal directional cut with the twist angle, $\theta=21.78^\circ$. If we consider the TBG flake with the full open boundary condition as shown in Fig.\,\ref{fig2}(c). Both the left and the right side edges are now gapped as the boundary is not consistent with $C_{2x}$ symmetry, but the effective mass gap closes at the $C_{2x}$ symmetric corners. As a result, the corner between the edges forms a domain wall. This domain wall of the gapped 1D edges effectively realizes the Su-Schrieffer-Heeger(SSH) chain\cite{PhysRevLett.42.1698}, where $e/2$ localized charge is placed on the top and the bottom corners. In result, the HOTI phase of the TBG enjoys the coincidence of the two fundamentally different topological invariants: second SW number protected by the inversion symmetry and the mirror winding number protected by $C_{2x}$ symmetry.

\section{Derivation of the edge Hamiltonian}\label{sec:edge}

In this section, we derive the Jackiw-Rebbi soliton model of the HOTI in Eq. \eqref{eq:H} directly from the low energy model of the twisted bilayer graphene. Our stating point is the low-energy model of the twisted bilayer graphene with the sublattice exchange even structure. The low energy Hamiltonian expanded near $K$ point can be written as\cite{PhysRevLett.114.226802},
\bea
\label{Eq:shtbg}
H_{tbg}=-v_F(-k_x\sigma_x\mu_z+k_y\sigma_y)+\gamma e^{i\theta \tau_z\sigma_z\mu_z}\tau_x e^{-i\theta \tau_z\sigma_z\mu_z}+V_0\tau_z,
\eea
where $v_F$ is the Fermi velocity. $V_0$ represents the external gate voltage. $\gamma$ and $\theta$ is the parameters that specifies the details of the interlayer coupling. $\sigma,\mu,$ and $\tau$ represent the Pauli matrices of the sublattice, the valley, and the layer degree of the freedom respectively. Focusing on the specific valley, say $\mu_z=1$, and $\theta=-\pi/2$, the following unitary matrix $U$, transforms Eq. \ref{Eq:shtbg} to the low energy model of the inversion symmetric HOTI proposed by Wang et al.\cite{wang2018higher}: 
\bea
U H_{tbg}U^\dagger =v_F(k_x \tau^x + k_y \tau^y \sigma^y )+\gamma \tau^z+V_{0}\tau^x\sigma^z,
\label{eq:Hbulk}
\\
U=\frac{1}{2}\left(
\begin{array}{cccc}
	e^{-\frac{i\pi}{4}   } & -e^{-\frac{i \pi}{4} } & -i e^{-\frac{i \pi}{4} } & -i e^{-\frac{i \pi}{4} } \\
	-i e^{-\frac{i \pi}{4} } & -i e^{-\frac{i \pi}{4}} & -e^{-\frac{i \pi}{4} } & e^{-\frac{i \pi}{4} } \\
	-i e^{\frac{i \pi }{4}} & i e^{\frac{i \pi }{4}} & e^{\frac{i \pi }{4}} & e^{\frac{i \pi }{4}} \\
	e^{\frac{i \pi }{4}} & e^{\frac{i \pi }{4}} & i e^{\frac{i \pi }{4}} & -i e^{\frac{i \pi }{4}} \\
\end{array}
\right).
\eea

The above Hamiltonian preserves the space-time inversion and the mirror symmetries, where each symmetry is explicitly defined as,
 \bea 
 T: H(\vec{r})=\tau^z H^*(\vec{r}) \tau^z,
 \\
 \nonumber
 P:
  H(\vec{r})=\tau^z H(-\vec{r}) \tau^z,
  \\
  \nonumber
  M:
  H(x,y)=\sigma^x H(x,-y)\sigma^x.
  \eea
It is important to note that the time-reversal and the inversion symmetry is different from those derived in the microscopic model of the TBG. This is because the symmetries are defined within a single valley. To derive the physical time-reversal and the inversion symmetries, we need to include the valley degree of the freedom. However, for the purpose of the Jackiw-Rebbi soliton construction, it is not necessary.
In addition, we consider the specific situation where the mass term forms a domain wall along a disk of radius $R$: $\gamma=|m|(\gamma=-|m|)$ for $r>R(r<R)$. The domain wall problem can be formulated by transforming the Hamiltonian into the polar coordinates as, 
\bea
H(r,\theta)=\gamma\tau^z-iv_F\Gamma^1(\theta)\partial_r-i\frac{v_F}{r}\Gamma^2(\theta)\partial_\theta,
\label{Eq:polarH} 
\eea
where $\Gamma_{1}(\theta)\mathord{=}\Gamma_{2}(\theta\mathord{-}\frac{\pi}{2})\mathord{=}\textrm{sin}(\theta)\tau^y\sigma^y \mathord{+}\textrm{cos}(\theta)\tau^x$. We find that the domain wall harbors a pair of counter-propagating edge modes circulating the disk: 
\bea
\Phi_{+,l}\mathord{\sim}e^{-|m|/v_F r+il\theta}
(e^{i\theta},\mathord{-}ie^{i\theta},i,1), \Phi_{-,l}\mathord{\sim}e^{-|m|/v_F r+il\theta} (-e^{-i\theta},-ie^{-i\theta},-i,1),
\eea
 where $l$ represents the angular momentum of the edge mode. We can check that the underlying symmetries acts in the edge modes as,
\bea
T: \Phi_{\pm,l}(\theta) \rightarrow -\Phi_{\mp,-l}(\theta),
\\
\nonumber
P: \Phi_{\pm,l}(\theta) \rightarrow -(-1)^l\Phi_{\pm,l}(\theta+\pi),
\\
\nonumber
M: \Phi_{\pm,l}(\theta) \rightarrow \pm i\Phi_{\mp,-l}(-\theta).
\eea

Unlike the conventional topological insulator problem, these edge modes are not protected by the underlying time-reversal and inversion symmetries. For example, we are allowed to add an angle dependent potential, $\Delta \tau^y\sigma^z \sin\theta$ to the Hamiltonian in Eq. \eqref{eq:Hbulk} without breaking the underlying symmetries. In the presence of the potential, the effective Hamiltonian can be written in the edge mode basis, $\phi_l(\theta)$=($\Phi_{+,l}(\theta),\Phi_{-,l}(\theta)$), as, 
\bea
H_{\textrm{eff}}(\theta)
= \left(
\begin{array}{cc}
	\frac{v_F}{R}(\frac{1}{2}-i\partial_\theta)   & \Delta \sin \theta  e^{i\theta} \\
	\Delta \sin \theta e^{-i\theta} & \frac{v_F}{R}(\frac{1}{2}+i\partial_\theta) \\
\end{array}
\right)
\label{Eq:inst}
\eea
The off-diagonal potential term in Eq. \eqref{Eq:inst} now gaps out the edge modes but vanishes at the angles $\theta\mathord{=}0$ and $\pi$. Due to this angular dependence, the localized corner states, $\Psi_0$ and $\Psi_\pi$, emerge at $\theta=0$ and $\theta=\pi$ respectively: 
\bea
\nonumber
\Psi_0(\theta \mathord{=}\epsilon) &=& \Psi_\pi (\theta \mathord{=}\pi\mathord{+}\epsilon) 
\\
&\mathord{\sim}& (e^{i\frac{1}{2}\epsilon-\frac{\Delta R}{v_F}\frac{\epsilon^2}{2}},-ie^{-i\frac{1}{2}\epsilon-\frac{\Delta R}{v_F}\frac{\epsilon^2}{2}}).
\label{Eq:corner}
\eea
This completes the Jakiw-Rebbi soliton construction of the inversion symmetric HOTI phase. It is important to note that the corner states have anti-periodic boundary condition. This is not an artifact of our approximation but indicating the presence of the intrinsic $\pi$ Berry phase. The Berry phase arises from the facts that the electron is embedded in a curved edge of higher dimensional bulk and it picks up phase of $\pi$ as it encircles around the edge\cite{PhysRevLett.105.156803,Cho2015}. 

After constructing the edge model, we consider the application of the external gate voltage, $V_{0}\tau^x\sigma^z$, which breaks the mirror symmetry. The matrix element between the edge state wave function can be calculated as, 
\bea
\langle \phi_l(\theta) |V_{0}\tau^x\sigma^z | \phi_{-l}(\theta) \rangle=0.
\eea
The gate voltage does not have local effect in the edge states, as each mirror sectors are spatially separated. However in a finite sized system, we can consider a hopping with different angles. For example, there exists a finite matrix element between the wave functions at $\theta$ and $-\theta$. The angular dependence of the matrix element is calculated as,
\bea
\langle \phi_l(\theta) |V_{0}\tau^x\sigma^z | \phi_{-l}(-\theta) \rangle=
V_0\left(
\begin{array}{cc}
	0 &  \sin \theta  \\
	- \sin \theta  & 0 \\
\end{array}
\right).
\label{Eq:s1}
\eea
We find that Eq. \eqref{Eq:s1} is equivalent to the gate voltage induced backscattering term in the main text:
\bea
H_{\textrm{gate}}(\theta)=V_0f(\theta)\Psi^\dagger(\theta) i\tau^y \Psi(-\theta),
\label{Eq:ftheta}
\eea
where $f(\theta)=sin(\theta)$.

\section{Derivation of path integral}\label{sec:path}

In this section, we derive the path integral of the edge Hamiltonian in Eq. \eqref{eq:H} in the main text. For the clarity, we start our discussion by writing Eq. \eqref{eq:H} again here.
\bea
H(\theta)=\psi^\dagger(\theta)\left(
\begin{array}{cc}
	i\frac{v_F}{R}\partial_\theta & \Delta \sin \theta  e^{i\theta}  \\
	\Delta \sin \theta  e^{-i\theta} & -i\frac{v_F}{R}\partial_\theta \\
\end{array}
\right)\psi(\theta)
\eea
To account the gate-voltage induced short-ranged backscattering, we enlarge the Hamiltonian into $4\times4$ matrix using the enlarged spinor $(
\Psi(\theta),\sigma_x\Psi(-\theta))$. The enlarged Hamiltonian is given as,
\bea
H_{total}(\theta)
=
\left(
\begin{array}{cccc}
	i\tilde{v}\partial_\theta & \text{$\Delta $}\sin \theta e^{i\theta}  & V (\theta)  & 0 \\
	\text{$\Delta $}\sin \theta e^{-i\theta}  & -i\tilde{v}\partial_\theta & 0 & -V (\theta)  \\
	V(\theta)  & 0 & i\tilde{v}\partial_\theta & -\text{$\Delta $}\sin \theta e^{i\theta}  \\
	0 & -V(\theta)  & -\text{$\Delta $}\sin \theta e^{-i\theta}  & -i\tilde{v}\partial_\theta \\
\end{array}
\right)
\eea
where $\tilde{v}=\frac{v_F}{R}$ and $V(\theta)=V_0f(\theta)$. The above Hamiltonian can be block-diagonalized into two independent sectors which is given as,
\bea
H(\theta)_\pm=\psi^\dagger_\pm(\theta)\left(
\begin{array}{cc}
	i\tilde{v}\partial_\theta \pm V(\theta) & \Delta \sin \theta  e^{-i\theta}  \\
	\Delta \sin \theta  e^{i\theta} & -i\tilde{v}\partial_\theta \pm V(\theta) \\
\end{array}
\right)\psi_\pm(\theta)
\eea
Since we doubled the Hamiltonian, the two sectors are not completely independent but they are related by the following condition: 
\bea
\psi_{+,1}(\theta)=\psi_{-,2}(-\theta), \psi_{-,1}(\theta)=-\psi_{+,2}(-\theta),
\label{eq:sboundary}
\eea
where $\psi_{\pm}=(\psi_{\pm,1},\psi_{\pm,2})$. We consider the following gauge transformation: $\psi_{\pm}(\theta)\rightarrow e^{\pm i\int^\theta_0 V(\theta)/\tilde{v} \sigma^z}\psi_{\pm}(\theta)$. The Hamiltonian transforms accordingly,
\bea
H(\theta)_\pm=\psi^\dagger_\pm(\theta)\left(
\begin{array}{cc}
	i\tilde{v}\partial_\theta  & \Delta \sin \theta  e^{- ig(\theta)}  \\
	\Delta \sin \theta  e^{i g(\theta)}  & -i\tilde{v}\partial_\theta  \\
\end{array}
\right)\psi_\pm(\theta)
\eea
where $g(\theta)=\theta \pm \frac{2}{\tilde{v}}\int_0^\theta d\theta'  V(\theta')$. Finally, we observe that the Hamiltonian of the two sectors can be thought as the pseudospin-$1/2$ system subject to the effective magnetic field rotating in the $x-y$ plane. Since the effective field of each sector rotates in the opposite directions, each sector acquires the opposite Berry phases. This is the source of the interference effect. It is important to note that the gauge transformation does not alter the identity between the two sectors in Eq. \eqref{eq:sboundary} if $V(\theta)=-V(-\theta)$, which is the condition that the gate voltage satisfies(See Eq. \ref{Eq:ftheta}).

%\begin{figure}[t!]
%	\centering\includegraphics[width=0.9\textwidth]{figs2}
%	\caption{(a) Schematic figure of the path integral procedure. There are two distinct paths connecting the corners states at $\theta=0$ and $\theta=\pi$ colored by the red and the blue line respectively. (b) We can decompose the two paths into two distinct sectors where the two sectors are subject to the magnetic fields with the same amplitude but winds in the opposite direction. This magnetic field configuration contributes to the opposite Berry phase between the two sectors, resulting in the interference effect.}
%	\label{figs2}
%\end{figure}

After presenting the basic idea behind the interference effect, we complete the path integral for the completeness. The above Hamiltonian is linear in momentum and the energy spectrum is unbounded. This feature forbids a simple Feynmann path integral of the action. To resolve this issue, we rather consider the squared Hamiltonian, $H_{\textrm{sq},\pm}=H_{\pm}^{2}$. $H_{\textrm{sq},\pm}$ is explicitly given as,
\bea
\nonumber
H_{\textrm{sq},\pm}&=&
\left(
\begin{array}{cc}
	(i\tilde{v}\partial_\theta)^2+\Delta^2\sin^2(\theta) & i\tilde{v}\Delta e^{ig(\theta)}(\cos(\theta)+i\sin(\theta)g')  \\
	-i\tilde{v}\Delta e^{-ig(\theta)}(\cos(\theta)-i\sin(\theta)g') & (i\tilde{v}\partial_\theta)^2 +\Delta^2\sin^2(\theta)\\
\end{array}
\right)
\\
&=&[(i\tilde{v}\partial_\theta )^2+\Delta^2\sin^2(\theta)]\sigma^0
+ i\tilde{v}\Delta ([ e^{ig(\theta)}(\cos(\theta)+i\sin(\theta)g')]\sigma_{+}+[-e^{-ig(\theta)}(\cos(\theta)-i\sin(\theta)g')] \sigma_-).
\nonumber
\\
\eea
Here, $\sigma_i$ represents the i-th Pauli matrices of the pseudospin and $\sigma_\pm=\frac{1}{2}(\sigma_x\pm i\sigma_y)$. The above Hamiltonian is now quadratic in momentum, which we can derive a well-regularized path integral. To derive the path integral form, we consider the imaginary time propagator,
\bea
\nonumber
\langle \theta_f, s_f | &e^{-\tau H_{sq,+}}&| \theta_i ,s_i \rangle
\\
&=&
\sum_{s_{1..n}=\pm}\int_{0}^{2\pi} \prod_{i={1..n}}d\theta_i \langle \theta_f, s_f | \theta_n ,s_n \rangle
\left[
\prod_{j=1}^{n-1}\langle \theta_{j+1} ,s_{j+1}| e^{-\epsilon H_{sq,+}}| \theta_j ,s_j \rangle
\right]
\langle \theta_1, s_1| \theta_i ,s_i \rangle
\label{Eq:path1}
\eea
where $| \theta ,s \rangle=|\theta \rangle \otimes |s \rangle$. $|s_{i,f}\rangle$ represents the initial and final pseudospin-polarized state.  Using the Baker-Campbell-Hausdorff formula, the matrix elements between the intermediate states can be decomposed into the two parts which are independent and dependent on the spin respectively:
\bea
\langle \theta_{j+1} ,s_{j+1}| e^{-\epsilon H_{sq,\pm}} | \theta_j ,s_j \rangle
\approx \langle s_{j+1}| \langle \theta_{j+1} | e^{-\epsilon H_0}
 e^{-\epsilon H_s}| s_j \rangle | \theta_j  \rangle,
\eea
where
\bea
\label{eq:hs2}
 H_0&=&[(i\tilde{v}\partial_\theta )^2+\Delta^2\sin^2(\theta)]\sigma^0,
 \\
 \nonumber
  H_s&=&i\tilde{v}\Delta ([ e^{ig(\theta)}(\cos(\theta)+i\sin(\theta)g')]\sigma_{+}+[-e^{-ig(\theta)}(\cos(\theta)-i\sin(\theta)g')] \sigma_-).
\eea
We first calculate the contribution of $H_0$ first. To do so, we first introduce the orthonormality condition of the periodic ring as,
\bea
\langle \theta | \theta' \rangle= \sum_{N=-\infty}^\infty \delta(\theta-\theta'+2\pi N) 
=\sum_{N=-\infty}^\infty \int_{-\infty}^{\infty} \frac{dp}{2\pi} e^{ i p(\theta-\theta'+2\pi N)} .
\eea
Using this normalization condition, we can further evaluate the spin independent matrix elements as,
\bea
\langle \theta_{j+1} &|& e^{-\epsilon [(i\tilde{v}\partial_\theta )^2 +\Delta^2\sin(\theta)^2]} | \theta_j  \rangle
\\
\nonumber
&\approx&
e^{-\epsilon(\Delta^2\sin^2(\theta_j))} ( 1-\epsilon(p\tilde{v})^2)  \sum_{N=-\infty}^\infty \int_{-\infty}^{\infty} \frac{dp}{2\pi} e^{ i p(\theta_{j+1}-\theta_j+2\pi N)} 
\\
\nonumber
&\sim& \sum_{N=-\infty}^\infty exp( -\frac{1}{4\epsilon \tilde{v}^2}(\theta_{j+1}-\theta_{j}+2\pi N)^2  -\epsilon\Delta^2\sin^2(\theta_j))).
\eea
By plugging in the above term to Eq. \eqref{Eq:path1}, We derive the following path integral,
\bea
\int_{\theta(\beta)=\theta_i}^{\theta(0)=\theta_f+2\pi N} &D\theta(\tau)& e^{ -\int_0^\beta d\tau \frac{\dot{\theta}^2 }{4 \tilde{v}^2} +\Delta^2\sin^2(\theta) }
\sum_{s_{1..n}=\pm} \langle s_f | s_n \rangle
\left[
\prod_{j=1}^{n-1}\langle s_{j+1}| e^{-\epsilon( B_{\textrm{eff}}\cdot \vec{\sigma} )}| s_j \rangle
\right]
\langle s_1 |s_i\rangle.
\eea
where $B_{\textrm{eff}}(\theta)$ is the effective magnetic field derived from the spin part of the Hamiltonian in Eq. \eqref{eq:hs2}.

We now need to evaluate the spin dependent part of the path integral. The spin part Hamiltonian can be replaced into the classical path of $\theta$ as,
\bea
\sum_{s_{1..n}=\pm} \langle s_f | s_n \rangle
\left[
\prod_{j=1}^{n-1}\langle s_{j+1}| e^{-\epsilon B_{\textrm{eff}}(\theta)\cdot \vec{\sigma}}| s_j \rangle
\right]
\langle s_1 |s_i\rangle
=
\langle s_f | T e^{-\beta B_{\textrm{eff}}(\theta)\cdot \vec{\sigma}}|s_i\rangle,
\eea
$T$ represents the imaginary time-ordering. We now apply the adiabatic approximation such that the spin fluctuation is ignored and the spin state follow the adiabatic evolution as the eigenstates of $ B_{\textrm{eff}}(\theta(\tau))$. 
\bea
\nonumber
\approx\sum_{\alpha=\pm}\langle s_f | B(\theta_f),\alpha \rangle \langle B({\theta_f}),\alpha | Te^{-\epsilon B_{\textrm{eff}}(\theta)\cdot \vec{\sigma}} | B(\theta_i) ,\alpha\rangle \langle B(\theta_i),\alpha|s_i\rangle
\\
=\sum_{\alpha=\pm}\langle s_f | \vec{B}(\theta_f),\alpha \rangle 
\langle \vec{B}({\theta_i}),\alpha|s_i\rangle
e^{- i \textrm{Im}\int_0^\beta \langle \vec{B}_{\textrm{eff}}(\theta(\tau)),\alpha |\partial_\tau | \vec{B}_{\textrm{eff}}(\theta(\tau)),\alpha \rangle 
+  |B_{eff}(\theta(\tau))|  },
\eea
where we have introduced the spin polarized state such that $\vec{B}\cdot \vec{\sigma}|\vec{B},\alpha\rangle=\alpha |\vec{B}||\vec{B},\alpha\rangle$.
Summing up all the contributions, we find the expression of the path integral as,
\bea
\langle \theta_f, s_f | &e^{-\beta H_{sq,+}}&| \theta_i ,s_i \rangle
=
\sum_{v=-\infty}^\infty\int_{\theta(\beta)=\theta_i}^{\theta(0)=\theta_f+2\pi N} D\theta(\tau) 
\\
\nonumber
&\times&
\sum_{\alpha=\pm}\langle s_f | \vec{B}(\theta_f),\alpha \rangle 
\langle \vec{B}({\theta_i}),\alpha |s_i\rangle
e^{ -\int_0^\beta d\tau \frac{\dot{\theta}^2 }{4 \tilde{v}^2} +\Delta^2\sin^2(\theta) } 
e^{- i \textrm{Im}\int_0^\beta \langle \vec{B}_{\textrm{eff}}(\theta(\tau)),\alpha |\partial_\tau | \vec{B}_{\textrm{eff}}(\theta(\tau)),\alpha \rangle 
	+  |B_{eff}(\theta(\tau))| }.
\eea
From Eq. \ref{Eq:corner}, we notice that the corner states at $\theta=0$ and $\theta=\pi$ have the same pseudo-spin polarization. Therefore, we can simplify the above expression as,
\bea
\langle \theta\mathord{=} \pi | e^{-H_{sq,+}\tau}|\theta\mathord{=}0\rangle
=
\sum_{v=-\infty}^\infty \int_{\theta(\beta)=\theta_i}^{\theta(0)=\theta_f+2\pi N} D\theta(\tau) e^{ -S_+} 
\eea
where the action is given as,
\bea
S_+=\int_0^\beta d\tau \frac{\dot{\theta}^2 }{4 \tilde{v}^2} +\Delta^2\sin^2(\theta) + |B_{eff}(\theta(\tau))|-i\textrm{Im} \langle \vec{B}(\theta(\tau)) |\partial_\tau | \vec{B}(\theta(\tau)) \rangle
\label{eq:s+}
\eea
Similarly, $H_{sq,-}$ sector has the Berry phase, but its value is opposite. Therefore, the Berry phase differences occur between the two sectors:
\bea
\Delta\gamma =\gamma_+-\gamma_-=2\gamma=\frac{4R}{v_F}\int_0^\pi d\theta V(\theta) 
\eea
Finally, we arrive at the result in Eq. \eqref{eq:gamma}.
\section{Derivation of tunneling amplitude}

Although the two corner states are inversion partners of one another, a intercorner tunneling, which lifts the degeneracy of the corner states, always present. In the presence of the tunneling, the eigenstates can be reconstructed by taking the linear combinations as,  $\Psi_\pm=\Psi_0 \pm \Psi_\pi$ with the energy splitting, $\Delta E=E_{+}-E_{-}$. The tunneling amplitude and the energy splitting can be formally calculated using the instanton method. We first notice that the real part of the action in Eq. \eqref{eq:s+} describes nothing more than a particle in a ring subject to a periodic potential\cite{PhysRevB.45.13544}. The classical equation of the motion of Eq. \eqref{eq:s+} is the sine Gordon equation, and it permits the following (anti-)instanton solution,
\bea 
\theta(\tau)_{I(A)}=2\arctan(e^{\pm 2\frac{v_F}{R}\Delta(\tau-\tau_0)}).
\eea

By plugging in the instanton solution, we derive the tunneling amplitude of the single instanton and anti-instanton respectively:
\bea
\lim_{\tau\rightarrow \infty}\langle \theta\mathord{=} \pi | e^{-H_{sq}\tau}|\theta\mathord{=}0\rangle_{I,A}
\mathord{=} K\tau e^{-\frac{\omega\tau}{2}} \sqrt{\frac{S_0\omega}{2\pi v_F}} e^{-S_0\pm i\gamma},
\nonumber
\\
\eea
%$S_0=\frac{\Delta R}{u}$ is the action of the instanton.
where $\omega=\Delta^2$ is the zero-point frequency of the sinusoidal potential. $K$ is the constant determinant, describing fluctuations from the saddle point(Please see ref. \cite{BAJNOK2000585,rajaraman1982solitons} for the explicit calculation). $S_0$ is the action of the single instanton.

The geometric phase leads to the interference effect between the instantons. To see this, we calculate the full tunneling amplitude consists of multiple instanton processes. Using the dilute gas approximation, we find the tunneling amplitude, which is given as,
\bea
&&\lim_{\tau\rightarrow \infty}\langle  \theta\mathord{=} \pi | e^{-H_{sq}\tau}|0\rangle
=\lim_{\tau\rightarrow \infty}\sum_n
\langle  \pi | n\rangle\langle n |0\rangle e^{-E_n\tau}
\nonumber
\\
&= &
\sqrt{\omega}e^{-\frac{\omega\tau}{2}}
\sum_{n_I,n_A\ge 0}^{n_I+n_A \textrm{odd}}
\frac{[K\tau\sqrt{\frac{S_0}{2\pi}}e^{-S_0}]^{n_I+n_A}[e^{i\gamma}]^{n_I-n_A}}{n_I!n_A!}
\nonumber
\\
&=&\sqrt{\omega}e^{-\frac{\omega\tau}{2}} \sinh (2K\tau \sqrt{\frac{S_0}{2\pi }}e^{-S_0} \cos(\gamma) ),
\label{eq:sinh}
\eea
where $n_I(A)$ represents the number of the instantons and the anti-instantons respectively. Since the initial state departs from $\theta=0$ and end up at $\theta=\pi$, we only counts the odd number of the instanton processes ($n_I+n_A$). Finally, comparing the left side and the right side of Eq. \ref{eq:sinh}, we derive the energy difference in Eq. \eqref{eq:dE} in the main text:
\bea
\Delta E^2 = 4K \sqrt{\frac{S_0}{2\pi }}e^{-S_0}| \cos(\gamma)| .
\eea

We find the oscillation of the energy splitting as a function of the geometric phase $\gamma$. It is important to note that this result is based on the global gauge structure arising from the intrinsic Berry phase, and do not depend on the details of the wave functions. 

\section{Methods of transport simulation}

The transport simulations are carried out using non-equilibrium Green function methods(NEGF) as implemented by Datta\cite{datta_2005}. The central device region consists of the corner states of the HOTI phase derived from the full tight-binding model. In addition, the four semi-infinite graphene nanoribbons are attached to the device region to model the transport contacts in Fig. \ref{Fig3} (a). The self energy of the contacts are calculated using the wide band limit. Finally, the Green function of the corner states are calculated as, 
\bea
G(\omega)=(\omega+i\eta -h_c-(\Sigma_1+\Sigma_2+\Sigma_3+\Sigma_4))^{-1}
\eea
where $\Sigma_{n}$ is the self-energy of n-th semi-infinite graphene nanoribbon. $\eta$ is an infinitesimal constant. $h_c$ is the truncated Hamiltonian of the corner states. Among the four contacts, the transmission from $n_i$-th contact to  $n_o$-th contact is calculated using the Landauer-B{\"u}ttiker formula\cite{datta_2005,Hirsbrunner_2019},
\bea
T_{n_i,n_o}(\omega)=\frac{2e^2}{h}Tr[G(\omega) \Gamma_{n_i} G^\dagger(\omega) \Gamma_{n_o}].
\eea 
where $\Gamma_{n} \equiv i(\Sigma_{n}-\Sigma_{n}^\dagger)$. In Fig. \ref{Fig3}, we calculate the transmission and the current by varying the gate potential of the central device region.

\end{widetext}

\end{document}